\documentclass[a4paper,12pt]{article}
\usepackage{epsfig}
\newcommand{\beq}{\begin{equation}}
\newcommand{\eeq}{\end{equation}}
\newcommand{\beqa}{\begin{eqnarray}}
\newcommand{\eeqa}{\end{eqnarray}}
\newcommand{\ba}{\begin{array}}
\newcommand{\ea}{\end{array}}

\newcommand{\hh}{\hat{H}}
\newcommand{\bp}{{\bf p}}
\newcommand{\bq}{{\bf q}}

\begin{document}

\begin{center}
\epsfig{file=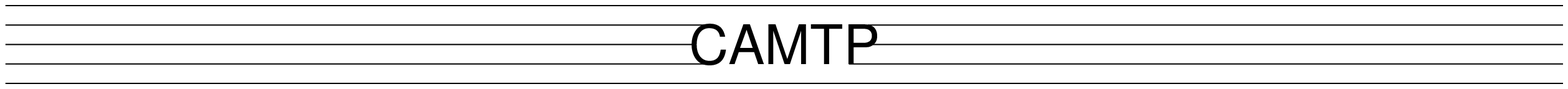,height=8mm,width=\textwidth}\\[2mm]
\end{center}

\begin{flushright}
Preprint CAMTP/98-6\\
June 1998\\
\end{flushright}

\vskip 0.5 truecm
\begin{center}
\Large
{\bf  Topics in quantum chaos of generic systems}\\
\vspace{0.25in}
\normalsize
Marko Robnik
\footnote{e--mail: robnik@uni-mb.si}\\
\vspace{0.3in}
Center for Applied Mathematics and Theoretical Physics,\\
University of Maribor, Krekova 2, SI-2000 Maribor, Slovenia\\

\end{center}

\vspace{0.3in}

\normalsize
\noindent
{\bf Abstract.} We review the main ideas and results in the
stationary problems of quantum chaos in generic (mixed) systems,
whose classical dynamics has regular (invariant tori) and
chaotic regions coexisting in the phase space. First we
discuss the universality classes of spectral fluctuations
(GOE/GUE for ergodic systems, and Poissonian for integrable
systems). We explain the problems in the
calculation of the invariant (Liouville) measure of
classically chaotic components, which has recently been
studied by Robnik et al (1997) and by Prosen and Robnik (1998).
Then we describe the Berry-Robnik (1984) picture,
which is claimed to become exact in the strict semiclassical
limit $\hbar\rightarrow 0$. 
However, at not sufficiently small values of $\hbar$ we see
a crossover regime due to the localization properties of
stationary quantum states where Brody-like behaviour with
the fractional power law level repulsion is observed in
the corresponding quantal energy spectra.

\vspace{0.6in}

PACS numbers: 03.65.-w, 03.65.Ge, 03.65.Sq, 05.45.+b, 05.40.+j, 
05.60.+w, 03.20.+i
\\\\
Published in {\bf Nonlinear Phenomena in Complex Systems} (Minsk, Belarus) 
Vol. 1, No. 1 (1998) 1-23
\normalsize
\vspace{0.1in}

\newpage

\section{Introduction}
The main problem of stationary quantum chaos is the following:
{\em Given a quantum Hamiltonian (operator) $\hh$ with infinitely
many bound states, being the quantized object based on its classical
analog $H=H({\bf q},{\bf p})$, the Hamiltonian $H$ as a function
of the $N$ coordinates ${\bf q}$ and momenta ${\bf p}$, what are
the geometrical and statistical properties of the eigenfunctions,
their Wigner functions, of the energy spectra and of the
matrix elements of other quantal observables.}
\\\\
Of course, we have the (stationary) Schr\"odinger equation,
which sometimes we can even solve analytically, e.g. in
cases of analytically solvable one-dimensional potentials or in cases 
of separable $N$-dimensional potentials (Landau and Lifshitz
1997), and sometimes we
can solve the underlying eigenvalue problem numerically,
at least in principle. However, the analytically solvable problems 
are very untypical, although quite important, because we can
use them to explore their neihbourhood (in the functional space
of Hamiltonians $H$) by means of a large variety of perturbational
techniques, and such a neighbourhood includes classically nonintegrable
systems which are typical (generic). This is in analogy of the
KAM systems in classical mechanics (Gutzwiller 1990). The numerical techniques
can be applied to almost all systems, but it turns out that
as soon as the system is not (classically) integrable and solvable, 
also the numerical techniques must be quite sophisticated,
especially if we ask for high lying eigenstates.
To know the Schr\"odinger equation and to have the potentiality
of solving it helps as little as the analogous situation in the
classical dynamics where the potentiality of solving the
Hamilton-Jacobi equation does not help very much in studying the
global, qualitative and quantitative, properties of motion
in generic nonintegrable Hamiltonian systems. This has been
realized for the first time by Henri Poincar\'e, who has 
shown (see Poincar\'e 1993, Goroff 1993) 
that the gravitational three-body problem is indeed
{\em nonintegrable} and this broken integrability can no longer
warrant the existence of invariant tori everywhere in phase
space, thereby giving the way to true chaotic motion, which 
cannot be embedded into smooth $N$-dimensional invariant surfaces.
The emergence of classically chaotic motion  gives rise to the
notion of qualitative dynamics, which is apt to embrace the richness and
variety of chaotic behaviour. To achieve that new methods
are needed, both analytically and numerically. Concepts
of Surface-of-Section (SOS) and similar ones become indispensible
to work in the new science of nonlinear dynamics, dealing with
the dynamical systems described by the set of $M$ ordinary
first order differential equations in their phase space.
For Hamiltonian systems we have $M=2N$ Hamilton equations.
On the quantum side we face a precisely analogous problem:
Given the (stationary) Schr\"odinger equation of a $N$-dimensional 
quantum system, whose classical analogue is not only nonseparable
but also nonintegrable and thus chaotic, we can hardly
see the structure of the solutions (eigenstates, described
by the wave function and the corresponding Wigner function)
and their global properties. New approach is necessary,
including the numerical one, to see and classify all the
possible types of behaviour. On the theoretical analytical
side the semiclassical methods are quite essential, and
this line of thoughts goes back to the pioneering and classical
work of Gutzwiller (Gutzwiller 1990 and references therein) 
and Percival (1973),
further developed by many workers in classical and quantum
chaos (Chirikov 1979, Casati and Chirikov 1994, Berry 1983,
Giannoni {\em et al} 1991, Haake 1991, Bohigas and Giannoni
1984, Bohigas 1991). For a recent excellent review, 
covering not only quantum chaos, but also all related theoretical
and experimental branches of physics see the paper by Weidenm\"uller 
and coworkers (Guhr {\em et al} 1998). 
\\\\
The purpose of this paper is to review the main methods and results
in the field of {\em quantum chaos}, i.e. the study of the solutions
of the Schr\"odinger equation connected with the classically 
nonintegrable and chaotic systems.

\section{The main assertion of stationary quantum chaos}

{\em The main assertion of stationary quantum chaos} is the following
answer to {\em the main problem of quantum chaos} in the semiclassical
limit of sufficiently small $\hbar$:

\subsection{Classical integrability}

The case ({\bf I}) of {\bf classically integrable quantal systems} $\hh$:
\\
If $H$ is classically integrable, then the wave function is locally
a superposition of a finite number of plane waves, the number of 
directions of the wave vector being equal to the number of possible
momentum vectors ${\bf p}$ through the coordinate point ${\bf q}$.
Globally, the probability density is equal to the classical probability
density obtained by projecting the $N$-dimensional invariant torus
onto the configuration space ${\bf q}$, up to within the resolution
scale of the order of one de Broglie wave length. The corresponding
Wigner function\footnote{For the definition and properties of Wigner
functions see section 3.}
 $W(\bq,\bp)$ of the eigenstate is a delta function
on the invariant torus labeled by the quantized classical action variable
${\bf I}(\bq,\bp) = {\bf I}_{\bf n}$, where ${\bf n}$ is the 
quantum number multi-index ${\bf n} = (n_1,n_2,\dots,n_N)$ denoting
the Maslov (EBK) quantized invariant $N$-torus, so that

\beq
W(\bq,\bp) = \frac{1}{(2\pi)^N} \delta_N( {\bf I}(\bq,\bp)-{\bf I}_{\bf n})
\label{eq:intWig}
\eeq
where $\delta_f$ is the $f$-dimensional Dirac delta function,
and in our case $f=N$.
The eigenvalues of $\hh$, i.e. the eigenenergies, in a small
interval, after unfolding, that is after reducing the mean energy
level spacing to unity,  obey (typically) the Poissonian statistics
(Robnik and Veble 1998):
The probability ${\rm E}(k,L)$ of observing $k$ levels inside an
interval of lentgh $L$ is given by\footnote{It can be shown that
knowledge of all ${\rm E}(k,L)$-statistics is equivalent to
the complete knowledge of all n-point correlation functions.
The calculation of ${\rm E}(k,L)$, however, is much easier
than the calculation of correlation functions etc., since
there is no binning and other advantages. This has been
pointed out very clearly by Aurich {\em et al} (1997). See
also the book by Mehta (1991).}

\beq
{\rm E}_{integrable} = {\rm E}_{Poissonian} (k,L) = \frac{L^k}{k!}\exp(-L)
\label{eq:EPoisson}
\eeq

The untypical cases have measure zero, and are characterized
by some number theoretic special properties like e.g. the rectangle
billiards with rational squared sides ratio.

\subsection{Classical ergodicity}

\noindent
The case ({\bf E}) of {\bf classically ergodic quantal systems} $\hh$:
\\
If $H$ is classically ergodic system, then the wave function is
locally a superposition of infinitely many plane waves, the directions
of the wave vector ${\bf k}$ being isotropically distributed on a 
$N$-dimensional sphere of
radius  $\hbar^{-1}\sqrt{2m(E-V(\bq))}$, if the Hamiltonian is 
$H(\bq,\bp) = \bp^2/(2m) + V(\bq)$, where $m$ is the mass and $V$
the potential energy. Due to the ergodicity the phases of the plane
waves are assumed to be random (Berry 1977a,b), which implies that the
wave amplitude $\psi(\bq)$ is a Gaussian random function. The random
phase assumption, however, can break down in vicinity of 
isolated unstable classical 
periodic orbits or families of such orbits, where
we observe {\em the scars} (Heller 1984), i.e. the regions of enhanced
probability density $|\psi|^2$. Thus, to the leading 
approximation we have the microcanonical Wigner function of
{\em almost all} eigenstates (Shnirelman 1979, Berry 1977a,b, Voros 1979),

\beq
W(\bq,\bp) = \frac{\delta_1(E- H(\bq,\bp))}
{\int d^N\bq d^N\bp \delta_1(E- H(\bq,\bp))}
\label{eq:Wigerg}
\eeq

The eigenenergies of $\hh$ in a small interval, after
unfolding, obey the predictions of classical {\em Random Matrix Theories}
(RMT), namely the statistics of the eigenvalues of the ensemble
of orthogonal Gaussian matrices (GOE) or of unitary Gaussian
ensembles (GUE) (depending on the existence or nonexistence
of an antiunitary symmetry), introduced by Wigner, and
also of the COE/CUE of Dyson (see e.g. the book by Haake (1991)). 
(We consider only systems
with classical analog $H$ and therefore ignore spin and GSE).
This assertion has been proposed by Bohigas, Giannoni and Schmit
(1984). It implies that ${\rm E(k,L)}$ statistics must obey the
RMT laws, the so-called BGS-Conjecture

\beq
{\rm E}_{ergodic} (k,L) =  {\rm E}_{RMT}(k,L)
\label{eq:ERMT}
\eeq

\subsection{Classically mixed systems}

\noindent
The case ({\bf M}) of {\bf classically mixed (generic) quantal systems} $\hh$:
\\
If $H$ is classically mixed system, then we can distinguish
between regular and irregular states.
Percival (1973) was the first to propose such a qualitative
characterization of eigenstates. The regular states are associated
with classical invariant tori (semiclassically EBK/Maslov quantized
tori, to the leading semiclassical approximation), and the chaotic
states are associated with chaotic components. This view has been
made more quantitative in the work of Berry and Robnik (1984).
The Berry-Robnik picture rests upon the 
{\em The Principle of Uniform Semiclassical Condensation} (PUSC,
see section 3),
which states that the Wigner functions of quantal states in
the limit $\hbar\rightarrow 0$ become positive definite,
and since they are mutually orthogonal, they must "live"
on disjoint supports, and the phase space volume (Liouville measure)
of each of them is of the order of $(2\pi\hbar)^N$. See section 3
and e.g. (Robnik 1997). The
question is, what is the geometry of the object on which
they "condense", and the answer - as a conjecture - is: 
uniformly on a classical invariant object (Berry 1977a,b, Robnik 1988,
1995, 1997).
Therefore we have regular and irregular states. The assumption is
that there is no correlation between the spectral
sequences  (regular and a series of 
irregular states). If $N\ge 3$ we have only one chaotic component
(the Arnold web of chaotic motion pervades the entire phase
space - energy surface - and is dense, i.e. its closure is the
energy surface)
and one associated irregular sequence of eigenstates, 
whereas in $N=2$ we have
many, even infinite number of sequences of irregular states, 
of smaller and smaller invariant measure,
each sequence being associated with one chaotic component.
It is thus assumed that the Wigner function of a regular state is
of type (\ref{eq:intWig}), whilst for irregular states, and generally,
it is

\beq
W(\bq,\bp) = \frac{\delta_f({\bf F}(\bq,\bp))\chi_{\omega}(\bq,\bp)}
{\int d^N\bq d^N\bp \delta_f({\bf F}(\bq,\bp))\chi_{\omega}(\bq,\bp)} 
\label{eq:mixWig}
\eeq
where $\chi_{\omega}(\bq,\bp)$ is the characteristic function of the
invariant component, labeled by $\omega$, 
being a (either smooth or nonsmooth, generally possibly also fractal) 
subset of the smooth $(2N-f)$-dimensional invariant
surface defined by the $f$ implicit equations (global integrals of 
motion), namely  ${\bf F}(\bq,\bp)=0$, where ${\bf F}=(F_1,F_2,\dots,F_f)$.
The characteristic function $\chi_{\omega}(\bq,\bp)$ is defined
to have value unity on $\omega$ and zero elsewhere.
The integer number $f$ can be anything between $1$ (ergodic system)
and $N$ (integrable system).
\\\\
Obviously, the formula (\ref{eq:mixWig}) is the most general
expression for a condensed Wigner function of a (pure)
eigenstate. It generalizes the cases ({\bf I}) and ({\bf E}).
Namely, if we have ergodicity, then $f=1$, 
we put $F_1(\bq,\bp)=E-H(\bq,\bp)$, and $\omega$ = entire energy surface,
and we recover equation (\ref{eq:Wigerg}). In the other extreme ({\bf I}),
we have $N$ global integrals of motion in involution, and
so  ${\bf F}(\bq,\bp) = {\bf I}_{\bf n} - {\bf I}(\bq,\bp)$,
and $\omega$ = the invariant torus labeled by ${\bf I}_{\bf n}$,
and we recover equation (\ref{eq:intWig}) of case ({\bf I}).
In the most general case, therefore, formula (\ref{eq:mixWig})
applies. Obviously, $W$ is normalized (see next section),

\beq
\int d^N\bq d^N\bp W(\bq,\bp) = 1
\label{eq:Wignorm}
\eeq
For generic (mixed) systems the most typical case is
$f=1$,  $F_1(\bq,\bp)=E - H(\bq,\bp)$ and $\omega$ is a
(nonsmooth, typically fractal, chaotic) subset of the energy surface
$F_1$. We write down this most important case explicitly:

\beq
W(\bq,\bp) = \frac{\delta_1(E-H(\bq,\bp))\chi_{\omega}(\bq,\bp)}
{\int d^N\bq d^N\bp \delta_1(E-H(\bq,\bp))\chi_{\omega}(\bq,\bp)} 
\label{eq:mixWigerg}
\eeq

It is important to know the relative invariant (Liouville) measure
of chaotic and regular eigenstates because the Hilbert space of
a mixed Hamiltonian system is split into regular and irregular
eigenstates, in the strict semiclassical limit, precisely
in proportion to the classical invariant measure of the
integrable component (invariant tori) and of the irregular components.

It is quite obvious by looking at the equation (\ref{eq:mixWigerg})
that the invariant Liouville measure of a subset $\omega$ of
the energy surface is equal to

\beq
\rho(\omega) = 
\frac{\int d^N\bq d^N\bp\delta_1(E-H(\bq,\bp))\chi_{\omega}(\bq,\bp)}
{\int d^N\bq d^N\bp \delta_1(E-H(\bq,\bp))} 
\label{eq:muomega}
\eeq
The relative invariant Liouville measure of the regular components
will be denoted by $\rho_1$, and the measures of chaotic
components (ordered in sequence of decreasing measure) by
$\rho_2,\rho_3,\dots,\rho_m$, where $m=\infty$ for $N=2$ and
$m=2$ for $N\ge 3$, as already explained. In section 4 I shall
explain how one can calculate the measures $\rho_2,\rho_3,\dots$.
\\\\
Assuming the above mentioned absence of correlations pairwise
between $m$ spectral sequences, due to the fact that they
have disjoint supports and thus do not interact, 
where $m$ is infinite for $N=2$
and $2$ for $N\ge 3$, the spectral statistics can be written as

\beq
{\rm E}_{mixed} (k,L) =  \sum_{k_1+k_2+\dots+k_m=k}
\prod_{j=1}^{m}{\rm E}_{j}(k_j,\rho_jL)
\label{eq:Emixed}
\eeq
which is a manifestation of Berry-Robnik (1984) picture.
Here ${\rm E}_j(k,L)$ is ${\rm E}_{Poisson}(k,L)$ for 
$j=1$, and ${\rm E}_{RMT}(k,L)$ for $j=2,3,\dots,m$.
See cases ({\bf I}) and ({\bf E}), equations (\ref{eq:EPoisson})
and (\ref{eq:ERMT}). The picture is based on the reasonable
assumption that (after unfolding) the mean density
of levels in the $j$-th sequence of levels is $\rho_j$,
simply applying the Thomas-Fermi rule of filling the
phase space volume with elementary cells of size
$(2\pi\hbar)^N$ in the thin energy shell embedding the
 corresponding subset $\omega$.
Therefore, please note that the second argument of ${\rm E}_j(k,L)$ is
weighted precisely by the classical relative invariant 
measure of the underlying invariant component. Also,
if there were several regular (Poissonian) sequences
they can be lumped together into a single Poissonian
sequence (which we traditionally label by $1$ with
relative invariant measure $\rho_1$): 
It is easy to show, that if $\alpha_1,\alpha_2,\dots,\alpha_l$ 
are positive real numbers and $\beta$ being their sum,
$\beta = \alpha_1+\alpha_2+\dots +\alpha_l$, then for all $k$, and $L$,
\beq
{\rm E}_{Poisson} (k,\beta L) =  \sum_{k_1+k_2+\dots+k_l=k}
{\rm E}_{Poisson}(k_1,\alpha_1 L){\rm E}_{Poisson}(k_2,\alpha_2 L)
\dots {\rm E}_{Poisson}(k_l,\alpha_l L)
\label{eq:Emixed2}
\eeq
by simply using the definition of ${\rm E}_{Poisson}(k,L)$ of
equation (\ref{eq:EPoisson}).  Thus, we have some kind of
a central limit theorem, saying that the statistically
independent superposition of Poisson sequences results
in a Poisson sequence, such that the total density of Poissonian levels
$\beta$ is equal to the sum of the partial level densities $\alpha_j$,
$j=1,2,\dots,l$. 
\\\\
The case ({\bf M}) is the most general one, and as the limiting
extreme cases includes cases ({\bf I}) and ({\bf E}).

\subsection{Limitations of the universality}

There are two important limitations of the above stated 
asymptotic behaviour as $\hbar \rightarrow 0$, when $\hbar$ is  not
yet small enough: One is the existence of the outer energy scale, 
and  the other one is the localization phenomena. 
As for the first, it has been shown by Berry (1985), applying the
semiclassical Gutzwiller periodic orbit theory (Gutzwiller 1990 and
the references therein),
that at energy scales (after unfolding!) $L \ge L_{max}$
we do not have the universality but typically a saturation,
i.e. ${\rm E}(k,L)$ statistics at $L$ larger than

\beq
L_{max} = \frac{\hbar}{T_0\langle \Delta E\rangle}
\label{eq:lmax}
\eeq
where $\langle \Delta E\rangle$ is the mean energy level 
spacing, and $T_0$ the period of the shortest classical
periodic orbit in the dynamical system $H(\bq,\bp)$,
deviate from their universal behaviour of
cases ({\bf I}) and ({\bf E}), equations (\ref{eq:EPoisson})
and (\ref{eq:ERMT}). Instead, e.g. the sigma and delta
statistics become constant. (For definition and inter-relationship
see section 3.) However, please observe that as 
$\hbar$ goes to zero, also $L_{max}$ goes to infinity
as a power $\hbar^{-N+1}$, so that in the semiclassical limit
the universality region becomes larger $\propto \hbar^{-N+1}$.
\\\\
As for the second limitation bordering the universal
behaviour we comment the following.
If the value of the (effective) Planck constant $\hbar$ is
not sufficiently small, then the eigenfunctions might not be
fully extended in the sense of the corresponding Wigner functions 
obeying the equations (\ref{eq:intWig}), (\ref{eq:Wigerg})
and (\ref{eq:mixWig}), but can be localized (not uniformly
extended) on the classical invariant object on which they
condense. Such a deviation from the ultimate limiting semiclassical
behaviour is therefore a manifestation of the localization
phenomena in stationary eigenstates of autonomous Hamiltonian
systems, and is manifested also in the spectral statistics.
For example, if we have a classically ergodic system, but with
very slow chaos (very large diffusion time), we shall observe
strongly localized states, mimicking a regular integrable system.
In such an extreme case of localization we shall observe Poissonian spectral
statistics rather than GOE/GUE. Depending on the strength
of localization we shall therefore be able to see transition
from Poissonian to GOE/GUE behaviour in an ergodic ssystem. 
In the intermediate
crossover regime we observe Brody-like behavior with fractional
power law level repulsion. In a KAM regime the same is true
if the effective $\hbar$ is not small enough to resolve
the structure of small chaotic components. We shall describe
these phenomena in the subsequent sections.
\\\\
However, just briefly, a qualitative comment is in order
at this place.
The relevant criterion for localization  is, that
the so called {\em break time} or {\em Heisenberg time},
defined through

\beq
t_{break} = t_{H} = \frac{2\pi \hbar}{ \langle \Delta E\rangle}
\label{eq:tbreak}
\eeq
where  $\langle \Delta E\rangle$ is the mean level spacing,
must be shorter than the diffusion time, so then we have strongly localized 
states, whilst in the opposite extreme we have strongly extended states.
The reason is very simple: Quite generally quantum mechanics
(of a suitably chosen initial wave packet) follows classical
dynamics (of a suitably chosen ensemble of initial conditions)
up to the break time, after which the interference phenomena set
in, resulting typically in destructive interference, and thus
in the stop of diffusion, which means localization (before
the entire phase space has been conquered). For example, in
two-dimensional billiards,  $\langle \Delta E\rangle$ is
just constant, so the break time is constant and independent
of energy, whilst the classical diffusion time, even if very large at
small energies $E$,  decreases
with energy as $const./\sqrt{E}$, so ultimately, as $E\rightarrow
\infty$, we shall find the extended states and then
the general picture of case ({\bf M}) is applicable,
which, of course, as the extreme cases, includes ({\bf I}) and ({\bf E}).
However, the phenomena of localization, including the scars,
are extremely important and as we have sketched above,
they are related to the important time scales which control
the finite time structure and behaviour of classical dynamics,
especially the transport times and so on.

\subsection{Distribution and fluctuation properties of
transition probabilities}

Finally, in this subsection  
we should comment on the statistical properties of
the matrix elements of other observables in the eigenbasis
of an integrable, ergodic and mixed system. The main work in
this direction has been done by Feingold and Peres (1986)
for the ergodic case, and this has been been generalized to integrable 
and mixed systems by Prosen and Robnik (1993a).
\\\\
The expectation values and generally
the matrix elements of other reasonable observables (Hermitian operators having
a classical limit) have been little studied (Feingold and Peres 1986,
Alhassid and Levine 1986, Wilkinson 1987, 1988). One well known result concerns 
the fluctuation properties of generalized 
intensities (squares of matrix elements)
within the framework of random matrix theories, namely the Porter-Thomas 
distribution (Brody {\em et al} 1981), which has been experimentally observed
and suggested by Porter and Thomas (1956) in the context of nuclear physics.
We expect that this fluctuation law applies also in classically ergodic
systems with few freedoms. The main motivation of our work (Prosen and 
Robnik 1993a) was to 
explain this and to find the appropriate generalization for Hamiltonian systems
in the transition region of mixed dynamics. 
\\\\
In order to study the fluctuation properties of generalized intensities one 
must be able to clearly separate the smooth mean part of the intensities
as the function of frequency (= energy difference between the
final and initial state/$\hbar$) from its fluctuating part. So, given the
frequency of the intensity we ask what is its mean value and which is the
distribution of its fluctuating part in units of the mean value. In the
classically ergodic case Feingold and Peres (1986) propose a formula 
expressing the mean intensities in terms of the power spectrum of the given
observable taken over a dense chaotic classical orbit. In deriving this result
they rely on the Shnirelman theorem (1979) expressing the quantum expectation
value of a reasonable operator as the classical microcanonical average. This
theorem is obvious once one has in mind that the Wigner distributions of the
eigenstates of a classically ergodic system in the semiclassical limit are 
just microcanonical distributions (Berry 1977a,b, Voros 1979), equation
(\ref{eq:Wigerg}). In order to
rederive Feingold-Peres formula and to generalize it we first point out that
the Shnirelman theorem applies also to the states in the regular and mixed 
regime if the classical average is taken over the relevant classical invariant
ergodic component, which supports the corresponding semiclassical eigenstate.
This can be an invariant torus, a chaotic component, or the entire
energy surface.
\\\\
Following Feingold and Peres (1986) we start by looking at the following sum
over eigenstates $k$ of eigenenergies $E_k$ for the transition elements
$A_{jk} = \langle j|\hat{A}| k\rangle$
\begin{eqnarray}
\sum\limits_k \exp\left(i(E_j-E_k)t/\hbar\right)|A_{jk}|^2 &=&
\sum\limits_k \langle j|e^{i E_j t/\hbar}\hat{A}|k\rangle \langle k|
e^{-i E_k t/\hbar}\hat{A}|j\rangle = \\ \nonumber
&=& \langle j|e^{i\hat{H}t/\hbar}\hat{A}e^{-i\hat{H}t/\hbar}\hat{A}|j\rangle =
\langle j|\hat{A}(t)\hat{A}(0)|j \rangle \label{eq:FP1}
\end{eqnarray}
Now we apply the generalized Shnirelman theorem, stating that in the
{\em semiclassical limit} this is equal to the classical average 
\begin{equation}
C_j(t) = \{A(t)A(0)\}_j
\end{equation}
over the invariant ergodic component labeled by $j$ which supports the 
semiclassical state $|j\rangle$. Using the ergodicity on the given invariant
component this two-point autocorrelation function can be expressed as the 
time average along a classical dense orbit (dense in the given invariant 
component, which e.g. can be an invariant torus, or a chaotic component,
or the entire energy surface)
\begin{equation}
C_j(t) = \lim_{T\rightarrow\infty}\frac{1}{T}\int\limits_{-T/2}^{T/2}
d\tau A(t+\tau)A(\tau).
\end{equation}
Next we replace the sum $\sum_k$ by the integral $\int dE_k \rho(E_k)$, where
$\rho(E)$ is the density of states, and perform the Fourier transform and obtain
\begin{equation}
\langle |A_{jk}|^2\rangle_j = \frac{S_j((E_k-E_j)/\hbar)}{2\pi\hbar\rho(E_k)}
\end{equation}
where the state $j$ is fixed and the average $\langle .\rangle_j$ is taken
over states $k$ within a thin energy shell of thickness of few mean level
spacings. Here 
\begin{equation}
S_j(\omega) = \int\limits_{-\infty}^\infty dt C(t) e^{i\omega t} = 
\lim\limits_{T\rightarrow\infty}
\frac{1}{T}\left\vert\int\limits_{-T/2}^{T/2} dt A(t) e^{i\omega t}\right\vert^2
\label{eq:PS}
\end{equation} 
is the power spectrum of a dense orbit in the invariant ergodic component $j$.
If $A$ has a nonvanishing mean value $\{A\}_j$ the $S_j(\omega)$ will have
a delta spike at $\omega=0$, and this can be removed by replacing $A$ in the
above formulas by $A-\{A\}_j$. To calculate the actual mean values of 
the intensities $|A_{jk}|^2$ we perform in the above formula (on the LHS)
also the averaging over the $j$ states microcanonically over the
thin energy shell around $E_j$ of sufficiently wide width such that
the corresponding semiclassical states uniformly cover the energy surface,
whilst on the RHS we correspondingly take the microcanonical average over
all initial conditions $j$ on the energy surface $E_j$. So the final
formula for the mean generalized intensities is 
\begin{equation}
\langle|A_{jk}|^2\rangle = \frac{\{S((E_j - E_k)/\hbar)\}_E}{2\pi\hbar\rho(E)}
\label{eq:GFP}
\end{equation}
By  $\{ .\}_E$ we denote the microcanonical average over the energy surface $E$.
The apparent asymmetry in $jk$ of this formula disappears in the semiclassical
limit $\hbar\rightarrow 0$. In the numerical evaluations described in
(Prosen and Robnik 1993a) we
applied the above formula with $\{S(\omega)\}_E$ and $\rho(E)$ being calculated
on the energy surface placed half way between $E_j$ and $E_k$, i.e. 
$E=(E_j + E_k)/2$. This choice is met to minimize the error at finite $\hbar$.
\\\\ 
Knowing the average value of intensities as a function of $\omega$ we can now
separate the smooth part from its fluctuating part by renormalizing the matrix
elements as follows
\begin{equation}
X_{jk} = \frac{A_{jk}}{\sqrt{\langle |A_{jk}|^2\rangle}}.
\end{equation}
The renormalized matrix elements $X_{jk}$ are now regarded as random variable
whose probability distribution is denoted by $D(X)$, which by definition has
unit dispersion, and naturally is expected to be even function of $X$,
$D(X) = D(-X)$, and so it has zero mean. In the classically ergodic case we
expect that quite generally the matrix elements of a given operator are
very well modelled by the GOE of random matrix theories (Brody {\em et al} 1981)
which predict Gaussian distribution for $D_{PT}(X)=\exp(-X^2/2)/\sqrt{2\pi}$,
which is equivalent to the so-called Porter-Thomas distribution for the
intensities $I=X^2$, namely $P(I)=\exp(-I/2)/\sqrt{2\pi I}$, see 
(Porter and Thomas 1956). In integrable cases one expects a vast abundance
of at least approximate selection rules which render most $X$ to become zero
implying that $D(X)$ approaches a delta function $\delta(X)$ in the 
semiclassical limit. This can be seen by considering matrix representation
of an operator in the basis of the torus quantized eigenstates of an
integrable system, as explained in detail in (Prosen and Robnik 1993b).
In the mixed type dynamics (KAM) in the transition region between integrability
and chaos we expect a continuous transition from $\delta(X)$ towards 
$D_{PT}(X)$.
More precisely, a semiclassical formula for $D(X)$ in such
transition region has been derived by Prosen (1994a)
by taking into account the fact that the only broadening
of $D(X)$ stems from the transitions between chaotic initial and chaotic
final states belonging to the same family of the invariant ergodic components
(continuously parametrized by the energy), while all
other transitions are almost forbidden. This work rests upon a 
more detailed analysis of higher autocorrelation functions and is reported 
on in (Prosen 1994b). More details and the numerical 
illustration of our results can be found in (Prosen and Robnik 1993a) and
in (Prosen 1994a,b).

\section{The Principle of Uniform Semiclassical Condensation and
more about the wave functions and statistics}

In this section we want to explain the main ingredients and
arguments leading to the equations (\ref{eq:intWig}),
(\ref{eq:Wigerg}) and (\ref{eq:mixWig}). The ideas go back
to Berry (1977a,b), Shnirelman (1979), Voros (1979), Robnik (1988, 1995,
1997).
To see that the quantum analogy of the stationary (aspects of) 
chaos works well it is necessary to look at the objects
uniquely determined by given eigenstates in such a manner
that one can compare the eigenstates to the classical
states (phase portraits at given energy). This can be achieved by
introducing the Wigner functions (transforms) of given eigenstates,
e.g. of the wave functions.
\\\\
With this procedure
we are building up a kind of the quantal phase space, in the
spirit of the Wigner-Weyl formalism (de Groot and Suttorp 1972),
in the following way: Let $\psi_n({\bf q})$ be the n-th wave function
(eigenfunction as a solution of the Schr\"odinger problem)
in the $N$-dimensional configuration space with ${\bf q}$ being a position
vector, then its corresponding Wigner function (or transform) 
is defined as:

\beq
W_n({\bf q,p}) = \frac{1}{(2\pi \hbar)^{N}} \int d^N{\bf X}
\exp(-\frac{i}{\hbar}{\bf p.X})
\psi_n({\bf q}-\frac{{\bf X}}{2})\psi_n^{\ast}({\bf q}+\frac{{\bf X}}{2})
\label{eq:Wigner}
\eeq
where $\ast$ denotes the complex conjugation. The Wigner function is
obviously real. However, unlike the classical 
distribution functions, the Wigner functions are 
{\em not} positive definite, which is a fundamental consequence
of the very nature of quantum mechanics: If they were positive
then the quantum mechanics would be identical to classical mechanics.
This is the essence of the Wigner theorem (about the quantal
phase space distributions).
On the other hand, the Wigner functions do have the correct property
that they become the configurational probability density $|\psi({\bf q})|^2$
when projected down onto the configuration space (i.e. integrating
(\ref{eq:Wigner}) over the momenta ${\bf p}$) and complementary,
they become the momentum probability density $|\varphi({\bf p})|^2$ 
when integrated over the entire configuration space, as can be
immediately verified from the definition ({\ref{eq:Wigner}).
Therefore $W_n$ integrates to 1 if $\psi_n$ is normalized to unity,
i.e.
\beq
\int W_n({\bf q,p}) d^N{\bf q}d^N{\bf p} = 1.
\label{eq:normalization}
\eeq
They also obey the orthogonality relation
\beq
(2\pi \hbar)^N \int d^N{\bf q}d^N{\bf p} W_n({\bf q,p})W_m({\bf q,p}) 
= \delta_{nm},
\label{eq:Orthogonality}
\eeq
($\delta_{nm}$ here is the Kronecker delta, i.e. discrete delta function,
equal to 1 if $n=m$ and zero otherwise)
which can be understood at once by recalling that $(2\pi\hbar)^NW_n$ is in fact
the Weyl symbol of the projection operator $P_n=|n><n|$. Further,
it can be easily seen that the absolute value of $W_n$ is bounded
from above, namely (Baker 1958, see also de Groot and Suttorp 1972,
and Berry 1977b)

\beq
|W_n({\bf q,p})|\le \left( \frac{1}{\pi\hbar} \right)^N
\label{eq:bound}
\eeq
showing that it can diverge only in the semiclassical limit
when $\hbar \rightarrow 0$. One can also see from (\ref{eq:Orthogonality})
that (when $n=m$) one has

\beq 
\int W_n^2({\bf q,p}) d^N{\bf q} d^N{\bf p} = 1/(2\pi\hbar)^N
\label{eq:square}
\eeq
Therefore unlike (\ref{eq:normalization}) the latter integral
(\ref{eq:square}) can diverge as $\hbar \rightarrow 0$, 
and this divergence can be due to large contributions for
large values of $|{\bf q}|$ and $|{\bf p}|$ or due to the
singularities of $W_n^2$ (Berry 1977a). The latter possibility
is the one that actually occurs, as shown by Baker (1958)

\beq
W_n({\bf q,p}) \rightarrow (2\pi\hbar)^N W_n^2({\bf q,p}), \;\;\;
as\;\;\; \hbar\rightarrow 0.
\label{eq:limitingW}
\eeq
Therefore in the semiclassical limit the Wigner functions $W_n$
become positive definite and divergent $\approx (2\pi \hbar)^{-N}$,
which is weak enough still to obey (\ref{eq:Orthogonality}) with
$n=m$. This orthogonality relation shows then that all pairs $W_n$,
$W_m$, with $n\not=m$ must have disjoint supports. Therefore they
are effectively nonzero and divergent only on a small piece of
volume of size $(2\pi\hbar)^N$. In fact this {semiclassical 
condensation} must take place close to and on the energy shell
around the classical energy surface of energy $E_n$, of total volume
$(2\pi\hbar)^N$. This is of course equivalent to the simple 
Thomas-Fermi rule of how to determine the average density of
states, semiclasically: Divide the available classical phase space
volume inside the given energy $E_n$ by $(2\pi\hbar)^N$, and on 
the average this must be equal to $n$, which is the sequential
quantum number, i.e. the (cumulative)  number of eigenstates
below the energy $E_n$. From these general considerations we
cannot conclude more than stated. It is not clear a priori
on what geometrical object does the Wigner function $W_n$ condense
as $\hbar\rightarrow 0$. Berry (1977a,b) and Voros (1979), in
agreement with Shnirelman's theorem (1979), have suggested
that in case of classical ergodicity $W_n$ condenses {\em uniformly}
on the energy surface, becoming the microcanonical distribution,
(\ref{eq:Wigerg}).
On the other hand, in the opposite extreme of an integrable
motion, or a KAM system with invariant tori, which are EBK
quantized and support the quantum state $|n>$, Berry (1977a)
derives from the semiclassical wave functions in coordinate
space that the corresponding Wigner function is equal to (\ref{eq:intWig}).
So here the Wigner function condenses uniformly on the EBK quantized torus.
\\\\
The latter result
can be easily obtained by noting that in the semiclassical
limit the classical canonical tranformations and the quantization
do commute, and thus can be performed directly in the
space of action-angle variables, immediately yielding the
above result, as shown in (Robnik 1995, Hasegawa {\em et al} 1989).
\\\\
In both extreme cases, the ergodicity (\ref{eq:Wigerg}), 
and the quasi-integrability (existence of a quantized KAM torus)
(\ref{eq:intWig}), we see that the Wigner function condenses
uniformly on the underlying classical object, which is
the invariant indecomposable component in the classical phase space.
It seems thus very natural to elevate these findings to the
{\em Principle of Uniform Semiclassical Condensation}, PUSC, (Robnik 1988,
Li and Robnik 1994, Robnik 1997), which claims the following: 
{\em In the semiclassical limit
$\hbar \rightarrow 0$ the Wigner function of the n-th eigenstate
condenses uniformly on the underlying classical invariant object
(topologically transitive component) labeled by $\omega$, 
which can be an invariant
$N$-torus, the entire energy surface (in case of ergodicity), 
or a chaotic component. The corresponding Wigner function is,
in the most general case, given by (\ref{eq:mixWig})}.
This principle has a great predictive power,
when accepted.  
\\\\
The statistical properties of the wave functions
in the coordinate space have been analyzed by Berry (1977b), where
he has shown that in ergodic cases the probability amplitude
distribution is Gaussian random functions, and its autocorrelation
function is described by the Bessel functions. In case of
integrability or quasi-integrability (a quantized invariant
KAM torus) the wave functions in coordinate space are 
quite ordered, they typically have caustics (projection
singularities when projecting the Wigner function (\ref{eq:intWig})
down the momentum space onto the configuration space). They are
locally well described by the {\em finite} number of plane waves,
because classically there is only a finite number of possible
trajectory velocities (obtained by projection of the torus and of the
quasiperiodic orbits on the torus onto the configuration space),
whereas in case of ergodicity the number of plane waves is
{\em infinite} (a circular ensemble of wave vectors), and they have uncorrelated
phases, which implies Gaussian randomness. See e.g. (Robnik 1988, 1995, 
1997).
The limiting behaviour of the condensing Wigner function on
a classical invariant object (no $\hbar$ enters in this equation,
so in a sense we have classical Wigner functions!) implies also
that the coarse grained probability density in the configuration
space is just classical probability density, obtained by projecting
the Wigner function onto the coordinate space, i.e. by integrating it
over the momenta $\bp$. By coarse grained
we mean smoothing over a few de Broglie wavelengths. And by the classical
probability density we mean the value proportional to the time
spent (asymptotically) in each cell of equal size (relative
invariant measure) in the discretized phase space. Recently,
we have brilliantly demonstrated this fact (Robnik {\em et al} 1998).
\\\\
>From the behaviour of the stationary wave functions we now turn back
to the study of the statistical properties of the energy spectra, based 
on PUSC, giving some more details than in section 2.
Again we restrict our discussion only to the time independent
(autonomous) Hamiltonian systems with finite (bounded) classical
motion and correspondingly a purely discrete quantal energy spectrum
(no scattering states). Further we asssume that there are infinitely
many energy levels, so that the questions of statistical properties
of energy spectra can be raised. Van Kampen (1985) has defined quantum chaos
as "{\em...that property that causes a quantum system to behave statistically".}
Now we have seen that this element is involved in the
morphology of the eigenstates of classically chaotic, especially
ergodic, systems. Therefore we must conclude that {\em stationary
quantum chaos} exists, and corresponds exactly to the classical
chaos. We shall now demonstrate that this is the case also when
studying the energy spectra (and possibly also the statistical
properties of other observables). 
\\\\
One of the most important cornerstones of the stationary quantum chaos
is the so-called {\em Bohigas-Giannoni-Schmit Conjecture} (1984),
BGS-Conjecture, introduced already in subsection 2.2,
equation (\ref{eq:ERMT}). 
It states that the classically ergodic Hamiltonian systems
(with discrete spectrum) exhibit universal spectral fluctuations,
whose statistical properties are correctly captured by the
conventional Random Matrix Theories (RMT) (Mehta 1991), 
and are thus universal. See case ({\bf E}) of section 2.
If we ignore spin (which is not important in studying the classical
limit) then the spectral fluctuations are described by either
the fluctuations of the eigenvalues in the  Gaussian Orthogonal Ensemble 
(GOE) of random matrices, if the system
has an antiunitary symmetry, or by Gaussian Unitary Ensemble (GUE),
if there is no antiunitary symmetry, such as e.g. time reversal 
symmetry, involved in the system. Of course, this statement applies to
the statistical analysis of the spectrum, after the
unfolding procedure, in which the actual physical energy is
replaced by the average number of states up to the given energy -
i.e. the integrated (cumulative) level distribution. After unfolding
the mean energy level spacing is by construction equal to unity. 
The average density of states typically is very well described
by the familiar Thomas-Fermi rule of filling the classical
phase space volume with the quantum cells of size $(2\pi\hbar)^N$,
for which we have seen the reason in the above analysis of
the semiclassical behaviour of the condensed Wigner functions
of eigenstates.
Sometimes the corrections to this {\em asymptotically exact} 
rule can be obtained, e.g. in plane and $N$-dimensional billiards,
constituting the famous Weyl rule with perimeter, curvature,
corner corrections etc (See e.g. Berry and Howls 1994 and
references therein). In this unfolding procedure the information 
on the (nonuniversal) average density of states is 
eliminated from the spectrum, giving way to the possibility
of {\em universal fluctuations of energy levels around its
nonuniversal mean distribution}. In the classically ergodic systems
this is exactly what we find, confirmed and supported by many
numerical and actual experiments, and theoretically first
corroborated by the result of Berry (1985) on the delta ($\Delta$) 
statistics, and recently claimed to be proven by Andreev {\em et al} (1996),
however, under much stronger conditions than ergodicity,
namely assuming the exponential decay of correlations.
For some recent review see (Robnik 1994,1995,1997).
An important recent work in this connection is by Keating {\em et al}
(1996,1997).
\\\\
What we said in the above paragraph applies to the scaling systems, where
the energy limit (of quantum number $n\rightarrow\infty$) is
somehow equivalent to the semiclassical limit of $\hbar\rightarrow 0$.
That is, there is a scaling variable involving the energy and $\hbar$ such 
that the classical dynamics is constant while energy is changing.
One such example are the billiard systems, among which the plane
billiards are most widely used models. Examples of rigorously
ergodic systems are the Sinai billiard, the stadium billiard of
Bunimovich and the cardioid billiard. The latter is  the limiting case of
the family of billiards with analytic boundaries defined as the 
quadratic conformal map of the unit disc, introduced by Robnik
(1983,1984) and further studied by many workers. See Robnik {\em et al}
(1997). 
In billiards the topology and the geometry of the phase portrait
are exactly identical at all energies except for the scaling of the momentum
as a square root of the energy. Therefore the limit of infinite
energy is equivalent to the semiclassical limit of $\hbar\rightarrow 0$.
The constancy of the classical dynamics across the spectral stretches
that we study is important to draw clear and safe conclusions 
about the relationship between the classical and quantum chaos.
If a system is not a scaling system, then there is no way out
other than taking just  a small energy interval and letting
$\hbar \rightarrow 0$ so that in this limit the interval is containing
an arbitrarily large number of energy levels, a necessary 
condition to introduce and to define statistical distributions.
This general case is the one that we assumed in section 2.
\\\\
There are two most important statistical measures used to characterize
the energy spectra, and both of them are easily related to
the ${\rm E}(k,L)$ statistics that we introduced in section 2. 
One is the {\em level spacing distribution}
usually denoted $P(S)$, where $S$ is the length of the spacing and
$P(S)dS$ is the probability that $S$ lies within the infinitesimal
interval $(S,S+dS)$. It is normalized to unit probability (by
definition of probability density) and to unit first moment
(due to the construction by the unfolding procedure). 
It can be shown that  $P(S)$ is the second derivative
of the so-called gap probability ${\rm E}(0,L)$ of having no
levels inside the interval of length $S=L$, 
so $P(S) = d^2 {\rm E}(0,S)/dS^2$.
$P(S)$ measures the short range correlations between energy levels.
Here the important point is the behaviour of $P(S)$ at small $S$:
For GOE one has the linear behaviour $P(S) \approx const\times S$
whilst for GUE we have $P(S) \approx const\times S^2$. Correspondingly
we talk about the {\em linear} and {\em quadratic level repulsion}:
Because $P(S) \rightarrow 0$ as $S\rightarrow 0$ the level crossings,
i.e. the degeneracies, are not likely, and in GUE this level
repulsion is stronger than in GOE, giving rise - paradoxically -
to a more regular spectrum. For GOE/GUE the corresponding $P(S)$
for the infinite dimensional case cannot be obtained in a closed form,
and for the details the reader is referred e.g. to (Bohigas 1991).
However, it is quite surprising and fortunate that the 2-dim GOE/GUE
models yield closed analytic formulae which give an excellent
approximation to the infinite dimensional case (which normally
is referred to when speaking about RMT results).
For GOE we have the so-called Wigner distribution

\beq 
P_{{\rm GOE}}(S) = \frac{\pi S}{2} \exp (-\frac{\pi S^2}{4}),
\label{eq:GOEWigner}
\eeq
and for the GUE case 

\beq
P_{{\rm GUE}}(S) = \frac{32 S^2}{\pi^2} \exp (-\frac{4 S^2}{\pi}).
\label{eq:GUEWigner}
\eeq
Both cases are easily derived by assuming the 2-dim Gaussian
real symmetric matrices (for GOE) or Hermitian symmetric 
matrices (for GUE). Interestingly, they can be derived from
the so-called Wigner surmise (see e.g. Bohigas and Giannoni 1984,
Brody {\em et al} 1981), which is an approximate argument outside
the scope of RMT; it is some kind of a statistical argument.
\\\\
Long range correlations are measured by the second most important
statistical measure, the delta statistics
$\Delta (L)$, introduced by Dyson and Mehta. It is an inverse measure
of spectral rigidity/regularity, as is immediately obvious from
the definition:

\beq
\Delta (L) = \langle min_{A,B} \frac{1}{L} \int_{-L/2}^{L/2}
\lbrack {\cal N}(x) - Ax-B \rbrack^2 dx \rangle
\label{eq:defdelta}
\eeq
where  ${\cal N}(x)$ is the {\em unfolded} cumulative spectral staircase
function  (${\cal N}(x) =$ linear average $x$ plus oscillatory part 
$\tilde {\cal N}(x)$), the minimum is
taken with respect to the parameters $A$ and $B$, and the average 
denoted by $\langle ... \rangle$ is taken over a suitable energy
interval over $x$. Thus from this very definition  $\Delta (L)$
is the local average least square deviation of the spectral staircase
${\cal N}(x)$ from the best fitting straight line over an energy
range $x$ of $L$ mean level spacings. The more regular the (unfolded) 
spectrum, the easier is to find a linear fit over $L$ levels, and
consequently the smaller is $\Delta (L)$. 
The fact that we try to find the best linear fit
to the spectral staircase implies that at small $L$ the delta
statistics $\Delta (L)$ always behaves as $L/15$ and thus carries
no information about the system at all. 
\\\\
Sometimes it is useful also to know the number variance, denoted
by $\Sigma^2(L)$, the dispersion of the number of levels  $n(L)$ 
in an interval  of length $L$, where $\langle n(L) \rangle = L$,
and

\beq
\Sigma^2(L) = \langle (n(L) - L)^2 \rangle = L - 2\int_0^L
(L-r) Y_2(r) dr
\label{eq:sigmastatistics}
\eeq
where  $Y_2(r)$ is the pair cluster function (Bohigas 1991,
Haake 1991, Mehta 1991). There exists also the connection
between $\Sigma^2$ and $\Delta$, 

\beq
\Delta (L) = \frac{2}{L^4} \int_0^L (L^3-2L^2r + r^3)\Sigma^2(r) dr
\label{eq:sigmatodelta}
\eeq
although, strictly speaking, this has been proven so far only within
the context of RMT (Aurich {\em et al} 1997, Mehta 1991).
Finally, as announced in section 2,
 we shall consider not the set of all the cluster functions
$Y_n(x_1,x_2,\dots, x_n)$, where $n=2,3,\dots$, but rather
the  $E(k,L)$ statistics, for all $k=0,1,2,\dots$, following the 
suggestion of Steiner and coworkers (Aurich {\em et al} 1997), because they are
very easy to calculate numerically and yet contain the complete
information about the spectral statistics. Since by definition
${\rm E}(k,L)$ is {\em the probability} that inside an interval
of length $L$ we find exactly $k$ levels, there are simple
relationships to other statistical quantities. For example,
as already mentioned, the level spacing distribution $P(S)$ is

\beq
P(S) = \frac{\partial^2}{\partial L^2} {\rm E}(k=0,L=S)
\label{eq:ezerotolsd}
\eeq
and

\beq
\Sigma^2(L) = \sum_{k=0}^{\infty} (k-L)^2 {\rm E}(k,L).
\label{eq:ekltosigma}
\eeq
and therefore, through (\ref{eq:sigmatodelta}), we have
the relation expressing $\Delta(L)$ in terms of the ${\rm E}(k,L)$ 
statistics.
\\\\
>From the definition of the
Poissonian ${\rm E}(k,L)$ statistics in (\ref{eq:EPoisson})
it is easily derived (see (\ref{eq:ezerotolsd}))

\beq
P_{Poisson}(S) = \exp (-S)
\label{eq:Poissonps}
\eeq
and after (\ref{eq:ekltosigma})
\beq
\Sigma^2_{Poisson}(L) = L,
\label{eq:Poissonsigma}
\eeq
and then using (\ref{eq:sigmatodelta})

\beq
\Delta_{Poisson} (L) = \frac{L}{15}.
\label{eq:Poissondelta}
\eeq
Poissonian statistics means also by definition that there
are no correlations, i.e. the pair correlation function
factorizes, so that  we have for the pair cluster function
(c.f. Mehta 1991, Bohigas 1991)

\beq
Y_2^{Poisson}(x) = 0.
\label{eq:PoissonY}
\eeq
Thus, using this fact in equation (\ref{eq:sigmastatistics}) and then
(\ref{eq:sigmatodelta}) we again recover Poissonian values 
(\ref{eq:Poissonsigma}) and (\ref{eq:Poissondelta}).
\\\\
It is at large $L$ that
the different universality classes of behaviour emerge. Interesting are
the asymptotic results for large $L$. For a completely
regular (equidistant) spectrum (like e.g. one-dimensional harmonic oscillator)
one obtains that for large $L \gg 1$ , $\Delta (L)$ is just constant
and equal to $1/12$. On the other hand the RMT gives for GOE

\beq 
\Delta_{{\rm GOE}} (L) \approx \frac{1}{\pi^2} \log L,
\label{eq:GOEDelta}
\eeq
and for GUE

\beq
\Delta_{{\rm GUE}} (L) \approx \frac{1}{2\pi^2} \log L,
\label{eq:GUEDelta}
\eeq
This result of RMT has been derived on dynamical grounds for
the energy spectra of {\em individual} classically ergodic systems, 
applying the Gutzwiller's
periodic orbit theory, in a remarkable and important paper by 
Berry (1985), giving some theoretical support to BGS-Conjecture.
As it is believed that the level repulsion is a purely
quantal effect and cannot be derived semiclassically (Robnik 1986,
1989),  one is not surprised that for eleven years there was hardly any
theoretical progress in establishing the BGS-Conjecture.
Indeed, Berry concluded (1991) that $P(S)$ cannot be derived
from applying periodic orbit theory, because it depends
sensitively on orbits of all lengths (periods). However,
recently Andreev {\em et al} (1996) claim to have derived
BGS-Conjecture by a different thinking, namely by studying the
spectrum of the Frobenius-Peron operator in the semiclassical limit,
using the techniques from the supersymmetry field theories,
especially the nonlinear sigma model (Weidenm\"uller {\em et al} 1985).
In considering the semiclassical limit they assume stronger
properties than ergodicity, namely the exponential decay
of (classical) correlations.
\\\\
We may conclude that the support for the BGS-Conjecture is
so strong that it can be regarded as well established, although
not rigorously proven as yet. By this I mean especially
the unusually strong and massive numerical support and
evidence accumulated during the past fourteen years.
Thus given the correctness of BGS-Conjecture we speak about
{\em the universality classes of spectral fluctuations},
namely the GOE and GUE class, of subsection 2.2. 
As is seen in the above formulae
for $P(S)$ and for $\Delta (L)$, and in the general equation
(\ref{eq:ERMT}), the universality is indeed
established: There is no parameter in the statistical properties
of spectral fluctuations, and the statistical measures are
identical for all ergodic systems, irrespective of their
dynamical and geometrical details. Turning this aspect around,
we conclude that the spectral fluctuations in a classically
ergodic system do not have any further information content.
When talking about BGS-Conjecture as applied to individual 
classically ergodic systems we must emphasize that if there
are any exact unitary symmetries involved in the system,
then they must be first eliminated before applying BGS-statement.
This process we call desymmetrization. Then, even after
desymmetrization, we still have to decide whether GOE or GUE
statistics apply: The general classification criterion is:
If the system has an antiunitary symmetry (like e.g. the 
time reversal symmetry) then GOE applies, and GUE otherwise
(i.e. if there is no antiunitary symmetry) (Robnik and Berry 1986,
Robnik 1986). 
\\\\
It is a surprise that RMT apply so well to {\em individual}
dynamical systems. We claimed that BGS-Conjecture holds true
in the strict semiclassical limit $\hbar\rightarrow 0$.
However, how small must be $\hbar$ to see this happening?
Thus we are now addressing the question of the limitations
to universality due to the not-sufficiently-small value
of the (effective) Planck constant.
\\\\
The following criterion is important: As soon as the
eigenstates  are fully extended chaotic (ergodic)
in the sense of (\ref{eq:Wigerg}) the BGS-Conjecture
applies. This is always happening asymptotically, as
$\hbar \rightarrow 0$, but a rough criterion is that the
classical diffusion time  (the typical time for the classical
dynamics to conquer the entire available phase space - energy
surface) is shorter than the break time introduced in section 2.4,
equation (\ref{eq:tbreak}):
If this inequality applies (strong enough, i.e. by a factor 
10 or so) then the eigenstates will be fully extended.
If the inequality is reversed, then the eigenstates are
chaotic (they lie in a classically chaotic region) but
{\em localized}, which means occupying only a small piece
(a proper subset) of the dynamically available phase space.
Obviously the break time  $(2\pi\hbar)/\langle\Delta E\rangle$ 
goes to infinity
as $\hbar \rightarrow 0$, whilst the classical transport time
is independent of $\hbar$, and thus the desired inequality
is asymptotically always satisfied, and therefore in the limit
all semiclassical states are fully extended states and we 
recover the universality of section 2.
\\\\
In such a dynamically localized classically ergodic regime
another interesting phenomenon occurs, namely the fractional
power law level repulsion, by which we mean that

\beq
P(S) \approx const \times S^{\beta},
\label{eq:fpllr}
\eeq
where the exponent $\beta$ can be $0$, $1$, $2$ or anything
in between. Thus the localization phenomena of the chaotic
eigenfunctions soften the strength of the level repulsion
($\beta$ going from $1$ to $0$). This phenomenon seems
quite obvious, since the tails of the localized wave functions
overlap even less and thus interact less strongly resulting
in reducing the value of $\beta$. This trend towards the
Poisson statistics is well known in the context of localization
phenomena in disordered solid state systems.  At present
we do not have a theory on how $\beta$ should be related
to the localization lengths/areas/volumes, and how to
predict them. But we have numerical examples  demonstrating
these features (see e.g. Prosen and Robnik 1994a,b). The
theory must satisfy the known limit of $\beta \rightarrow 1$
when $\hbar \rightarrow 0$. One distribution function
which at present does not have any deep physical justification as yet,
but is just a nice mathematical model which captures the
global level spacing distribution with the local property
(\ref{eq:fpllr}) is the well known Brody distribution (Brody 1973,
Brody {\em et al} 1981)

\beq
P_{{\rm Brody}}(S;\beta) = 
a S^{\beta} \exp (-bS^{\beta+1}), \;\;\; a=b(\beta +1),
\;\;\; b=\{\Gamma(\frac{\beta+2}{\beta+1})\}^{\beta+1}
\label{eq:Brody}
\eeq
where $a$ and $b$ are obviously determined by the normalizations of
the total probability and of the first moment to unity.
For $\beta =0$ we have Poisson distribution (exponential, i.e.
$P(S)=\exp (-S)$), and for $\beta =1$ we have Wigner (i.e.
2-dimensional GOE, given in equation (\ref{eq:GOEWigner})).
The role of dynamical localization phenomena has been first
realized by Chirikov {\em et al} (1981) in the time-dependent
systems like kicked rotator and Rydberg atoms in microwaves,
but has been also suggested in the time-independent Hamiltonian
systems (Chirikov 1993). Feingold (1996) has recently found
deeper relationship between the two phenomena.
\\\\
After having explained the two universality classes of spectral
fluctuations in the classically ergodic systems, we now have
to add and rediscuss the third universal class comprising of
classically integrable systems of two or more degrees of freedom
(Robnik and Veble 1998), see subsection 2.
(Systems with only one degree of freedom are exceptional
and special in the sense that as $\hbar \rightarrow 0$ 
the local spectrum is just the perfectly regular equidistant
spectrum.) Indeed, if we have two or more quantum numbers the
entire spectrum can be thought of as being composed of
an infinite number of statistically uncorrelated number
sequences, which of course must result in the Poisson
statistics ${\rm E}(k,L)$ in (\ref{eq:EPoisson}), and
specifically  (\ref{eq:Poissonps}), (\ref{eq:Poissonsigma}) 
and (\ref{eq:Poissondelta}).
\\\\
There are semiclassical arguments resting upon the torus 
quantization by Berry and Tabor (1977) showing that the 
statistics should be Poissonian. The torus quantization
(EBK quantization) is embodied in equation (\ref{eq:intWig}),
describing the associated Wigner functions.
However, these semiclassical
arguments are only approximation and it is far from obvious that
they should correctly describe e.g. the fine structure of energy
spectra and thus the level repulsion and their absence,
as has been recently pointed out by Prosen and Robnik (1993c,d).
In fact, as explained above, it is believed that semiclassics
cannot explain the level repulsion (short range correlations),
since it is a purely quantum effect.
We know that there are exceptions from the Poissonian
behaviour which have been rigorously proven to exist and
involve some highly nontrivial and sophisticated mathematics
(Bleher {\em et al} 1993). Nevertheless, there is quite massive
numerical and experimental support to the statement that
the spectral fluctuations of classically integrable systems
are quite accurately described by the Poisson statistics
(\ref{eq:EPoisson}) (Robnik and Veble 1998). 
There might be cases where the
statement is rigorous, whereas in general we think that
the measure of exceptions is small and maybe vanishing
in some sense. In every case this delicate problem persists
to be very important and difficult, but it should also be
accepted that typically Poisson model is an excellent
approximation. The most important feature is the absence of
short range correlations implying the absence of
level repulsion, which means that degeneracies are allowed
and this is mathematically exhibited in $P(S) \rightarrow const\not=0$,
in fact according to (\ref{eq:Poissonps}) we have $P(S)\rightarrow 1$
as $S\rightarrow 0$.
\\\\
Another remark should be made about a limitation to
universality, mentioned in subsection 2.4, the behaviour
of $\Delta (L)$ in individual dynamical systems at large $L$,
where the limitations to universality of section 2.4 set in. 
As discovered
by Casati {\em et al} (1985) and later explained by Berry (1985)
there is the phenomenon of saturation, by which we mean that
$\Delta (L)$ becomes effectively constant and equal $\Delta_{\infty}$
if $L > L_{{\rm max}}$, 
in {\em any} system (ergodic, integrable, partially chaotic - KAM type).
This leveling off of the delta statistics is nonuniversal,
but the $L_{max}$ can be estimated in the context of Berry's
theory (1985) as in equation (\ref{eq:lmax}).
Therefore for any $N\ge 2$
the onset of saturation $L_{max}$ goes to infinity
in the semiclassical limit  $\hbar \rightarrow 0$,
giving way to the full universality of the three universality
classes (\ref{eq:GOEDelta}, \ref{eq:GUEDelta}, \ref{eq:Poissondelta}).
The details of the saturation value $\Delta_{\infty}$
at a fixed and nonzero $\hbar$ can be found in (Berry 1985).
\\\\
Finally, we should say something more about the $P(S)$ and
$\Delta (L)$ in the classically mixed systems, thus
giving more details of the Berry-Robnik (1984) picture. 
Both statistics are of course
implied by the most general statistics (\ref{eq:Emixed}), through
the formulae (\ref{eq:ezerotolsd}) for $P(S)$, and through
(\ref{eq:ekltosigma}) and (\ref{eq:sigmatodelta})
for sigma and delta statistics.
\\\\
Using the general equation (\ref{eq:Emixed}) and approximation
(\ref{eq:GOEWigner}) to first calculate ${\rm E}_{GOE}$,
and then to find $P(S)$ through (\ref{eq:ezerotolsd}),
for $m$ level sequences,
we obtain an explicit analytic formula for $P(S)$, namely

\beq
P_m(S) = \frac{d^2}{dS^2} \lbrack \exp (-\rho_1S) 
\prod_{j=2}^{m} {\rm erfc} (\frac{\sqrt{\pi}}{2} \rho_jS)\rbrack
\label{eq:mBerry-Robnik}
\eeq
where ${\rm erfc}(x) = \frac{2}{\sqrt{\pi}}\int_{x}^{\infty} dt \exp(-t^2)$
is the complementary error function. (For GUE one has to use
(\ref{eq:GUEWigner}) instead of (\ref{eq:GOEWigner}).)
One can show

\beq
P_m(S=0) = 1 - \sum_{j=2}^{m} \rho_j^2,
\label{eq:Pzero}
\eeq
so that now as a consequence of the statistically indpendent
superposition of a number of level (sub)sequences we get a trend
towards Poissonian statistics, since the degeneracies become
possible due to the lack of level repulsion among the
regular levels on the one hand and among the levels
belonging to different sequences. Thus, e.g. even if there is
no regular component but two equally strong chaotic components,
so that $\rho_1=0$, but $\rho_2=\rho_3=1/2$, we get
$P_m(S=0) = 1 -1/4 =3/4 \not =0$. Only if there is only one
chaotic (ergodic) component we find $P_m(S=0) =0$, describing
the GOE-like level repulsion. 
\\\\
Most important in practical applications is the 2-component
Berry-Robnik formula ($m=2$), namely

\beq
P_2(S,\rho_1) = \rho_1^2 \exp(-\rho_1S) {\rm erfc} (\frac{\sqrt{\pi}}
{2} \rho_2 S) + (2\rho_1\rho_2+\frac{1}{2} \pi \rho_2^3S)
\exp (-\rho_1S -\frac{1}{4} \pi \rho_2^2S^2),
\label{eq:Berry-Robnik2}
\eeq
and we see the special case of equation (\ref{eq:Pzero})

\beq
P_2(S=0,\rho_1) = 1 - \rho_2^2 = \rho_1(2-\rho_1),
\label{eq:Pzero2}
\eeq
which vanishes only iff $\rho_1=0$ and $\rho_2=1$ (ergodicity).
This level spacing distribution is very important especially
in practical applications, because in mixed systems typically
we have a very large dominant chaotic region, so that the next
largest chaotic region is much smaller by orders of magnitude,
say only one percent of the leading one and can be neglected. 
In such case the two component formula  (\ref{eq:Berry-Robnik2}) 
is an excellent approximation.
\\\\
For the delta statistics (\ref{eq:defdelta}) one can derive
{\em the additivity property} implied by the statistical independence
of the (sub)sequences. First one shows it for the sigma statistics 
(the number variance) (see e.g. Bohigas 1984,1991). 
Then one can show (Seligman and Verbaarschot 1985)

\beq
\Delta (L) = \sum_{j=1}^{m} \Delta_j(\rho_jL).
\label{eq:Delta-m}
\eeq
\\\\
It is now interesting to verify whether Berry-Robnik theory applies
to actual systems which we can analyze numerically. We (Prosen
and Robnik 1993c,1994a,b)  have done such analysis for a certain 
one-parameter family of billiards, introduced by Robnik (1983, 1984),
namely the 2-dim billiard shape defined as the complex quadratic conformal
mapping  $w=w(z)=z + \lambda z^2$ of the unit disc  $|z|\le 1$
in the $z$-plane onto the $w$-plane. The boundary curve is known
in the theory of analytic curves as the Pascal's Snail. (Usually
workers refer to this billiard system as Robnik billiard, because
it was introduced and dynamically analyzed by the author.) 
The system is very important because it has analytic boundaries 
up to the limiting value of the shape parameter $\lambda=1/2$
where the singularity appears at $z=-1$, and therefore for small
$\lambda$ the KAM-Theory applies: At $\lambda=0$ we have the
integrable case of the circle billiard with conserved angular
momentum. For small $\lambda \le 1/4$ we have a convex billiard
with analytic boundaries, studied extensively especially by
Lazutkin (1981,1991), where much can be said about the classical
and semiclassical analysis, including a construction of an
approximate integral of motion (Robnik and Berry 1985, Robnik 1986).
This is essentially KAM scenario. For $\lambda > 1/4$ the boundary
is nonconvex and the KAM theory does not apply because the bounce
map becomes discontinuous.
When $\lambda=1/4$ the first point of zero curvature appears
at $z=-1$ and according to Mather (1982)  this guarantees that
all the Lazutkin caustics (generated by invariant tori for
glancing orbits supporting the whispering gallery modes of
quantal eigenstates in the semiclassical picture) are destroyed,
giving way (preparing the way) for ergodicity, which has been
postulated by Robnik (1983). In fact, a careful analysis of
certain periodic orbits (Hayli {\em et al} 1987) 
has shown that also for $\lambda> 1/4$
they can be stable, surrounded by very tiny stability islands 
that can hardly be detected numerically, which was the
reason why in the early work (Robnik 1983) they have not been
seen. According to their estimates the system has stable
islands up to $\lambda\approx0.2791$, whilst Li and Robnik (1996)
have numerical evidence that ergodicity is possible for
$\lambda \ge 0.2775$.
Recently it has been rigorously proven by Markarian (1993) 
that for $\lambda =1/2$ (we have a cusp singularity at $z=-1$,
because $dw/dz=0$ there) the system is ergodic, mixing and K.
This is thus the first billiard system with analytic boundaries
having the chance to be ergodic for $0.2775 < \lambda \le 1/2$.
(The Sinai billiard and the stadium of Bunimovich are rigorously
ergodic, mixing and K, but they do not have analytic boundaries.)
The system has been recently studied by many workers (Berry and
Robnik 1986, Robnik and Berry 1986, Frisk 1990, Bruus and Stone 1994,
Stone and Bruus 1993a,b, B\"acker {\em et al} 1995, Bruus
and Whelan 1996).
\\\\
The system is thus ideal to study the morphology of quantum eigenstates
at various $\lambda$, following a continuous transition from the
domain of torus states and Poisson statistics, through the regime
of the generic behaviour of mixed dynamics, to the extreme case of
(rigorous) ergodicity and entirely chaotic states with GOE statistics. 
(Of course, as explained in the introduction, we must separate
the exact symmetry classes of a given dynamical system before
performing the statistical analysis of the energy spectra. This
procedure is called the desymmetrization. In our case we have even
and odd reflection symmetry classes.)  The main results on
this have been published in (Li and Robnik 1995a,b,c)
\\\\
For the details please see (Robnik 1997) and the references
therein. Another brilliant
numerical confirmation of the Berry-Robnik statistics was
given in (Prosen and Robnik 1994a,b) for the quantized compactified
standard map, and by Prosen (1995,1996) for a semiseparable
oscillator, and recently for a quartic billiard (Prosen 1998).
The main problems and issues cocerning the Berry-Robnik picture
have been expounded recently in  our comment (Robnik and Prosen 1997). 
The deviation
from Berry-Robnik regime towards the Brody-like behaviour
with fractional power law level repulsion, as described in
subsection 2.4, has been analyzed in detail in (Prosen and Robnik
1993c, 1994a,b).

\section{Statistical properties of classically chaotic motion and
the measure of chaotic components}

In this section  we address the question of the statistics of
classically chaotic  motion and the problem of how to determine
the measure of the chaotic components $\rho_2, \rho_3,\dots$,
which in this section we shall simply denote by $\rho_2$,
relevant for the problem of stationary quantum chaos in mixed systems,
dealt with in subsection 2.3.
\\\\
In a recent work (Robnik {\em et al} 1997) we have demonstrated
some general scaling laws in the behaviour of stochastic diffusion
in strongly chaotic systems (ergodic, mixing and K with large
Lyapunov coefficient, i.e. large KS entropy), mainly in
Hamiltonian systems, or in the strange attractors of
dissipative systems. The so-called {\em random
model} that we developed describes very well the diffusion
on chaotic components, in the sense that the relative
(invariant) measure  $\rho(j)$ as a function of
the discrete time\footnote{We work either with mappings or with
Poincar\'e mappings on the surface of section. In each case $j$
is the number of the iterations of the map.} 
$j$ approaches unity exponentially as 

\beq
\rho(j) = 1 - \exp (-j/N_c)
\label{eq:exp}
\eeq
where $N_c$ is the number of cells of equal size (relative invariant measure)
$a=1/N_c$ into which the whole ergodic component is decomposed,
provided $N_c$ is sufficiently large, say  $N_c>100$ or so.
In the above equation we have defined $\rho(j) = \rho_2(j)/\rho_2(\infty)$.
Thus the average measure of occupied domain on the grid
of cells is $\langle ka \rangle = \rho (j)$.
This {\em random model} rests upon the assumption that there
are absolutely no correlations, not even between two consecutive
steps, so that at each step (of filling the $N_c$ cells) we have
the same {\em a priori probability} $a=1/N_c$ of visiting any of
the cells, irrespective of whether they are already occupied or not.
Such absence of correlations can be implied and expected by the large
Lyapunov exponents, which in turn imply strong stretching and
folding (of a phase space element) even after one iteration,
meaning that such a phase element will be evenly distributed
(in the coarse grained sense) over the entire phase space (or surface
of section). The universal scaling property is reflected in the
fact that $\rho(j)$ is a function of the ratio $(j/N_c)$ only,
and does not depend on $j$ and $N_c$ separately.
\\\\
Such an assumption of absence of all correlations
appears to be strong at first sight, and therefore
it is quite surprising that the model describes a whole
lot of deterministic dynamical systems for which we can expect
large Lyapunov exponents, namely 2D billiard (Robnik 1983,
$\lambda=0.375$), 3D billiard (Prosen 1997a,b, $a=-\frac{1}{5}$, 
$b=-\frac{12}{5}$), ergodic logistic map (tent map), 
hydrogen atom in a strong magnetic field ($\epsilon= -0.05$)
(Robnik 1981, 1982, Hasegawa {\em et al} 1989), and
standard map at ($k=400$), in which the agreement is almost perfect,
except for the last two systems where we see some long-time
deviations  on very small scales. However, in the standard map at
$k=3$, and in H\'enon-Heiles (1964) system at $E= \frac{1}{6}$ the
deviations are noticeable  though not very big (about only 
1\%). 
\\\\
It is also quite astonishing that the random model
applies very well even to {\em ergodic-only} systems, with 
strictly zero Lyapunov exponents, namely in case of  the
rectangle billiards (Artuso {\em et al} 1997), where the deviations
from the exponential law (\ref{eq:exp})
on the largest scale is within a few percent only.
It is a well known result (Sinai 1976) that polygonal billiards
have exactly zero Lypunov exponents, easy to understand since
all periodic orbits are marginally stable (parabolic), and since
they are everywhere dense, we conclude that the Lyapunov 
exponents must be zero everywhere.
\\\\ 
As a small but interesting comment we should mention our
results on testing the random number generators from (Press {\em
et al} 1986), where two of them (ran0 and ran 3) are found to be in 
perfect agreement with the random model, whilst the other two
(ran1 and ran2) are exhibiting big deviations. Thus, indeed,
some deterministic dynamical systems like hydrogen atom in
strong magnetic field etc. can be better number generators
than some built-in (black-box) computer algorithms. It should
be acknowledged, however, that there are other random number
generators which pass all tests of randomness, including
ours, e.g. in (Finocchiaro {\em et al} 1993).
\\\\
The random model developed in (Robnik {\em et al} 1997) is a statistical model
which predicts not only the average relative measure of occupied
cells  $\rho(j)=\langle ka \rangle$, the average taken over $k$, 
in (\ref{eq:exp}), but also the standard deviation $\sigma (j)$,
which under the same assumption of sufficiently large $N_c$
is equal to, to the leading order,

\beq
\sigma (j) = \sqrt{ \langle (ka)^2 \rangle - \langle (ka) \rangle^2 }
= \sqrt{ \frac{1-\rho(j)}{N_c} },
\label{eq:sig}
\eeq
and gives us an estimate of the size of expected statistical
fluctuations in $\rho (j)$. 
\\\\
The random model (Robnik {\em et al} 1997) 
has been subsequently generalized in an
important direction (Prosen and Robnik 1998), 
namely to describe the diffusion on
chaotic components in systems of mixed dynamics, with
divided phase space, having regular regions (invariant tori)
coexisting in the phase space with chaotic regions, a typical
KAM scenario (Kolmogorov 1954, Arnold 1963, Moser 1962,
Benettin {\em et al} 1984, Gutzwiller 1990). Such systems
in two degrees of freedom can have the fractal boundary between
the regular and irregular component and thus the convergence
to the theoretically expected results can be very slow, mimicking
a departure from the random model, although ultimately it
conforms to this model. In three or more degrees of
freedom there is no boundary between the regular and chaotic regions,
because we have the Arnold web (Chirikov 1979), which is everywhere dense
in the phase space, and thus a naive box-counting would imply
always that the relative invariant measure of the
chaotic component is equal to the measure of the entire 
phase space, so $\rho_2(j) =1$, which is wrong, because 
the KAM theorem gives rigorously that the relative measure
of the regular component $\rho_1$ is strictly positive,
$\rho_1 > 1$, moreover  it is close to unity with the
perturbation parameter. We assume that the invariant measure
of the chaotic component is positive, although strictly speaking
this is a major open theoretical problem in the mathematics
of nonlinear systems, the so-called coexistence problem (Strelcyn 1991).
Therefore in such case one must
introduce the possibility of different {\em a priori probabilities},
which now are no longer just the same and equal to $a=1/N_c$,
but have a certain distribution described by the so-called
greyness distribution $w(g)$, where $g$ is a continuous variable
on the interval $[0,1]$: $g=0$ means no visits (white cells), 
$g=1$ are
the most frequently occupied cells (black cells), and those
cells with $0 < g < 1$ have intermediate number of visits
(grey cells). With this model
we have shown  how by measuring (numerically calculating)
$w(g)$ we can determine the relative invariant measure $\mu$
of the chaotic component. The result is

\beq
\mu = \int_0^1 g w(g) dg
\label{eq:mu}
\eeq
and the time dependent relative measure of occupied
domain is equal to

\beq
\rho (j) = 1 - \int_0^1 dg w(g) \exp (-\frac{gj}{\mu N_c})
\label{eq:gexp}
\eeq
and the standard deviation is still given precisely by the 
equation (\ref{eq:sig}). The greyness distribution can be 
calculated numerically quite easily by noticing that the 
greyness $g$ is proportional to the average occupancy
number $n(g)$, namely $n(g) = g/\mu$, so by measuring
$n(g)$ in the limit $j\rightarrow \infty$ and after
normalizing the $g$ of the peak of $n(g)$ to unity, we get the $g$'s,
and then by binning them into bins
of suitably small size $\Delta g$ we get the histogram for $w(g)$.
\\\\
In case of ergodicity (only one
chaotic component) we have $w(g) = \delta (g-1)$, the 
Dirac delta function at $g=1$, and then from equations
(\ref{eq:mu}) and (\ref{eq:gexp}) follows  the random 
model, with exponential behavior (\ref{eq:exp}).
\\\\
In a later work we dealt with only {\em ergodic} systems, but such
having several components, each of them also being ergodic,
but weakly coupled, by which we mean that the transition
probability for going from one to another component is
very small and the typical transition time $j^{\ast}$
very long. Obviously, at small times we shall find the
random model (\ref{eq:exp}) with $N_c$ being equal to the
number of cells of the starting component, $N_c=N_1$,
whilst for very large times $j$, bigger than the typical
transition time $j^{\ast}$, so $j \gg j^{\ast}$, we shall
find again the random model  (\ref{eq:exp}), but now
with $N_c$ being equal to the number of all cells in
the system, $N_c=N_s$. In between, when $j\approx j^{\ast}$,
we have the crossover regime which we analyse in the present work.
So, the finite time structure of ergodic systems controlled
by their transport times can be also captured analytically.
We call the corresponding model the multi-component random
model of diffusion. For more details see (Robnik {\em et al} 1998).
\\\\
As the concluding remark of this section it should be explained
that the relative measures $\rho_1,\rho_2,\dots,\rho_m$ 
we need in the formula for the statistics of the classically 
mixed systems in subsection 2.3, equation (\ref{eq:Emixed}),
are the relative Liouville measures {\em on the energy surface} ($E$),
the surface being defined by $H(\bq,\bp)=E=const.$, 
whereas in this section we
explained how to calculate the relative (invariant simplectic) 
measure $\mu$ of a given invariant chaotic set $\omega$ 
{\em on the Surface of Section} (SOS). They are not the same, 
for a given chaotic set, but the procedures to calculate them are
relatively simply  related to each other, as has been
explained  by Meyer (1985).  The answer is
obtained from the general relation

\beq
\langle A \rangle_{E} = \int_{E} d^N\bq d^N\bp A(\bq,\bp)
\delta_1(E-H(\bq,\bp)) =
\langle \tau A\rangle_{SOS} = \int_{SOS} dX \tau A(\bq,\bp),
\label{eq:fromSOStoE}
\eeq
where $dX$ is the invariant simplectic measure on SOS,
saying, that the microcanonical average of any classical function $A(\bq,\bp)$
(observable) over the energy surface $E$ is equal to the average
over the SOS of $A$ weighted by the average time of recurrence $\tau$ of a 
trajectory to SOS on each invariant (ergodic) component. $\tau$ 
is constant for a given trajectory and thus also inside a given
invariant (ergodic) component, but changes from one component
to another, also on irrational invariant tori.
Thus if, specifically, $A_{\omega}(\bq,\bp)$ is the
characteristic function on a chaotic invariant 
component, so equal to $\chi_{\omega}(\bq,\bp)$,
then we have for the measure $\rho(\omega)$ of equation
(\ref{eq:muomega})

\beqa
\rho(\omega) & = &\frac{\int_Ed^N\bq d^N\bp \delta_1(E-H(\bq,\bp))
\chi_{\omega}(\bq,\bp)}{\int_Ed^N\bq d^N\bp \delta_1(E-H(\bq,\bp))} 
\nonumber \\
             & = &
\frac{\int_{SOS} dX \tau \chi_{\omega}(\bq,\bp)}
{\int_{SOS} dX \tau} = 
\frac{\tau_{\omega} \int_{SOS} dX \chi_{\omega}(\bq,\bp)}
{\int_{SOS} dX \tau} 
\label{eq:rho2frommu}
\eeqa
The third equality in the above equation 
follows by recalling that $\tau$ is constant on
a given invariant component $\omega$, equal to $\tau_{\omega}$.
Of course, we can and should calculate (numerically) $\rho(\omega)$
by discretizing the SOS into a network of $N_c$ cells
of equal size $a=1/N_c$. As is evident from equation (\ref{eq:mu}),
the invariant measure $dX$ in a given cell can be chosen simply 
equal to

\beq
dX = gw(g) dg
\label{eq:dX}
\eeq
This solves our problem of determining the classical measures
$\rho(\omega)$ in the context of their role in the semiclassical
limiting behaviour of the eigenstates and of the Hilbert space of
a general, classically mixed, and therefore generic system, 
dealt with in the context of quantum chaos mainly in subsection 2.3.

\section{Discussion and conclusions}

The main purpose of this paper is first to provide a compact
review of the main topics in the stationary quantum chaos 
in the general quantal Hamiltonian systems with discrete 
energy spectra, in correspondence with classical chaos,
in the strict semiclassical limit where the effective
Planck constant $\hbar$ is sufficiently small. This
problem has been expounded in the Introduction, section 1.
\\\\
In section 2, 
we have explained the universality classes of spectral fluctuations
and of their wave functions and of Wigner functions.
These comprise of the classically integrable systems
exhibiting Poisson statistics (subsection 2.1), and of the
classically  ergodic  systems, obeying the statistics of the
of the eigenvalues of the 
ensembles of random matrices from the classical Random Matrix
Theories, namely GOE and GUE, depending on whether the
system has or not an antiunitary symmetry, like e.g.
the time reversal symmetry (subsection 2.2). 
Then we went on explaining 
the general case of classically mixed dynamics, in the
transition region between classical integrability and
ergodicity, very often well described by the scenario
of the KAM Theory (subsection 2.3). 
We have shown how the Berry-Robnik (1984)
picture can be applied, and have generalized it to
arbitrary statistics, specifically ${\rm E}(k,L)$
statistics. The main message is that the Hilbert space
of eigenstates of a mixed system is decomposed into
the set of regular states (associated with a EBK/Maslov
quantized invariant torus) and a set if irregular
(chaotic) eigenstates, where their properties are
captured by the most general semiclassical Wigner
function in equation (\ref{eq:mixWig}). The
regular and irregular sequences are split exactly
in proportion to the relative invariant Liouville
measures of their supporting classical invariant sets.
Then we have described some limitations
to the universality classes and the semiclassical
asymptotical behaviour (subsection 2.4). The first
limitation stems from the existence of the Berry's outer
energy scale (after unfolding) $L_{max}=(2\pi\hbar)/
\langle \Delta E \rangle$ (see equation (\ref{eq:lmax})),
so that at $L\ge L_{max}$ we have no longer universality
but saturation (e.g. of the sigma and of delta statistics).
Nevertheless, as $\hbar\rightarrow 0$, $L_{max}$ goes to
infinity. The second limitation comes from the
localization phenomena, controlled by the relation of
the two time scales, the break time and the classical diffusion
time. If break time for a given stationary eiegnstate is 
shorter than diffusion (ergodic) time, then we have a strong
localization. If the system is ergodic (but slowly diffusing),
then we find a departure from the universal RMT statistics, and
in extreme case can be close to Poisson statistics. In
the opposite extreme, we find the extended states and
correct behaviour generally described by the mixed case
({\bf M}) of subsection 2.3. In subsection 2.5 we give
some fundamental results on the statistical properties
of general matrix elements, generalizing the important
Feingold-Peres (1986) theory.
\\\\
In section 3 we discuss the fundamental propeties of
the Wigner functions of the eigenstates, and introduce
the {\em Principle of Uniform Semiclassical Condensation}.
Then we say more about the wave functions and the 
spectral statistics, especially about the level
spacing distributions, sigma and delta statistics.
There we have also recalled Van Kampen's definition of
quantum chaos and concluded that the quantum chaos,
as a phenomenon "{\em ...causing the quantum systems
to behave statistically...}", does exist in the
problem of the stationary Schr\"odinger equation and
its solutions. 
\\\\
In section 4 we deal with the problem of how to determine,
also numerically, the relative invariant measure of 
classically chaotic
components, which enter in the semiclassical formulae
for statistics of mixed systems (\ref{eq:Emixed}).
We have described the main ideas, the approach and the
results of our recent works on this subject, giving the
final decription of how to proceed. The main goal 
of the theory of stationary quantum chaos is to
explain and to describe the mixed systems. Then, the
quantal parameters $\rho_j$ determined by analyzing
the spectral statistics must be equal to the purely
classical relative invariant (Liouville) measures
(subsection 2.3).
\\\\
We believe that the material presented here is also a
stimulation for further theoretical and numerical work,
and also shows the need for more rigorous results on the 
side of the mathematical physics. For example, we
still need the proof of the BGS-Conjecture, especially in
its full generality presented in equation (\ref{eq:ERMT}).
We need to prove the Principle of Uniform Semiclassical
Condensation, which is more general than the BGS-Conjecture.
Finally, as one of the most important issues, we need
to understand more deeply and quantitatively the
localization phenomena in the stationary eigenstates
of general systems. Some important new steps in this 
direction have been recently undertaken by Casati and
Prosen (1998a,b), and by Krylov and Robnik (1998).
 The most general aspects of quantum
chaos related to other branches of theoretical and
experimental physics have been recently reviewed and discussed in
detail by Weidenm\"uller and coworkers (Guhr {\em et al}
1998).

\section*{Acknowledgments}

I wish to thank the Editors of {\em Nonlinear Phenomena
in Complex Systems},  Professors V.I. Kuvshinov and
V.A. Gaisyonok for the kind invitation to
write and to submit this paper for the first issue
of this new international journal.
This work was supported by the Ministry of Science and
Technology of the Republic of Slovenia, and by the
Rector's Fund of the University of Maribor.
I also thank Dr. Toma\v z Prosen (University of Ljubljana)
for collaboration on many subjects discussed and
presented in this paper.

\newpage

\section*{References} 
\parindent=0. pt

Alhassid Y and Levine R D 1986 {\em Phys. Rev. Lett.} {\bf 57} 2879
\\\\
Andreev A V, Agam O, Simons B D and Altshuler B L 1996 {\em Phys. rev. 
Lett.} {\bf 76} 3947
\\\\
Arnold V I 1963 {\em Usp. Mat. Nauk SSSR} {\bf 18} 13
\\\\
Artuso R, Casati G and Guarneri I 1997 {\em Phys. Rev. E} {\bf 55} 6384
\\\\
Aurich R, B\"acker A and Steiner F 1997 {\em Int. J. Mod. Phys.} {\bf 11} 
805
\\\\
B\"acker A, Steiner F and Stifter P 1995 {\em Phys. Rev.} {\bf E 52} 2463
\\\\
Baker G A Jr. 1958 {\em Phys. Rev.} {\bf 109} 2198
\\\\
Benettin G C, Galgani L, Giorgilli A and Strelcyn J.-M. 1984
{\em Nuovo Cimento B} {\bf 79} 201
\\\\
Berry M V 1977a {\em J. Phys. A: Math. Gen.} {\bf 10} 2083
\\\\
Berry M V 1977b {\em Phil. Trans. Roy. Soc. London} {\bf 287} 30
\\\\
Berry M V 1983 in {\em Chaotic Behaviour of Deterministic Systems}
eds. G Iooss, R H G Helleman and R Stora  (Amsterdam: North-Holland) 
pp171-271
\\\\
Berry M V 1985 {\em Proc. Roy. Soc. Lond. A} {\bf 400} 229
\\\\
Berry M V 1991 in {\em Chaos and Quantum Physics} eds.
M-J Giannoni, A Voros and J Zinn-Justin (Amsterdam: North-Holland) 
pp251-303
\\\\
Berry M V and Robnik M 1984 {\em J. Phys. A: Math. Gen.} {\bf 17} 2413
\\\\
Berry M V and Howls C J 1994 {\em Proc. Roy. Soc. Lond. A} {\bf 447} 
527
\\\\
Berry M V and Robnik M 1986 {\em J. Phys. A: Math. Gen.} {\bf 19} 649
\\\\
Berry M V and Tabor M 1977 {\em Proc. Roy. Soc. Lond. A} {\bf 356} 375
\\\\
Bleher P M, Cheng Z, Dyson F J and Lebowitz J L 1993 {\em Commun. Math. 
Phys.} {\bf 154} 433-469
\\\\
Bohigas O 1991 in {\em Chaos and Quantum Physics} eds.
M-J Giannoni, A Voros and J Zinn-Justin (Amsterdam: North-Holland) pp87-199
\\\\
Bohigas O and Giannoni M-J 1984 {\em Lecture Notes in Physics} {\bf 209} 1
\\\\
Bohigas O, Giannoni M.-J. and Schmit C 1984 {\em Phys. Rev. Lett.} {\bf 
25} 1
\\\\
Brody T A 1973 {\em Lett. Nuovo Cimento} {\bf 7} 482
\\\\
Brody T A, Flores J, French J B, Mello P A, Pandey A and Wong S S M 1981
{\em Rev. Mod. Phys.} {\bf 53} 385
\\\\
Bruus H and Stone A D 1994 {\em Phys. Rev.} {\bf B 50} 18 275
\\\\
Bruus H and Whelan N 1996 CATS/NORDITA Preprint Copenhagen
\\\\
Casati G and Chirikov B V 1994 in {\em Quantum Chaos: Between Order and
Disorder} eds. G. Casati and B.V. Chirikov (Cambridge: Cambridge University
Press)
\\\\
Casati G, Chirikov B V and Guarneri I 1985 {\em Phys. Rev. Lett.} {\bf 54}  
1350
\\\\
Casati G and Prosen T 1998a {\em The quantum mechanics of chaotic 
billiards}, Preprint, to be published in {\em Physica D}
\\\\
Casati G and Prosen T 1998a {\em Quantum localization and cantori
in chaotic billiards}, Preprint, to be published in {\em Phys. Rev. Lett.}
\\\\
Chirikov B V 1979 {\em Phys. Rep.} {\bf 52} 263
\\\\
Chirikov B V, Izrailev F M and Shepelyansky D L 1981 {\em Sov. Sci. Rev. 
C2} 209
\\\\
Chirikov B V 1993 private communication
\\\\
Feingold M 1996 {\em Loalization in Strongly Chaotic Systems} Preprint 
Beer-Sheva
\\\\
Feingold M and Peres A 1986 Phys. Rev. A {\bf 34} 591
\\\\
Finocchiaro P, Agodi C, Alba R, Bellia G, Coniglione R, Del Zoppo A,
Maiolino C, Migneco E, Piattelli P and Sapienza P 
1993 {\em Nucl. Instrum. Methods.} {\bf 334} 504 
\\\\
Frisk H 1990 Preprint NORDITA -- 90/46 S
\\\\
M-J Giannoni, A Voros and J Zinn-Justin 1991 
{\em Chaos and Quantum Physics} eds. (Amsterdam: North-Holland) 
\\\\
Goroff D L 1993 {\em New Methods of Celestial Mechanics} (American 
Institute of Physics)
\\\\
de Groot S R and Suttorp L G 1972 {\em Foundations of Electrodynamics}
(Amsterdam: North Holland)
\\\\
Guhr T, M\"uller-Groeling A and Weidenm\"uller H A 1998,
{\em Random Matrix Theories in Quantum Physics: Common Concepts},
MPI Preprint  {\bf H V27} 1997, MPIfK Heidelberg, 
(cond-mat/9707301, 29 July 1997)  {\em Phys.Rep.} in press
\\\\
Gutzwiller M C 1990 {\em Chaos in Classical and Quantum Mechanics}
(New York: Springer)
\\\\
Hasegawa H, Robnik M and Wunner G 1989 {\em Prog. Theor. Phys.
Supplement (Kyoto)} {\bf 98} 198
\\\\
Haake F 1991 {\em Quantum Signatures of Chaos} (Berlin: Springer)
\\\\
Hayli A, Dumont T, Moulin-Ollagier J and Strelcyn J M 1987 {\em J. Phys. 
A: Math. Gen.} {\bf 20} 3237
\\\\
Heller E J 1984 {\em Phys. Rev. Lett.} {\bf 53} 1515 
\\\\
H\'enon M and Heiles C 1964 {\em Astron. J} {\bf 69} 73
\\\\
Keating J and Bogomolny E 1996 {\em Phys. Rev. Lett.} {\bf 77} 1472
\\\\
Keating J P and Robbins J M 1997 {\em J. Phys. A: Math. Gen.} {\bf 30} L177
\\\\
Kolmogorov A N 1954 {\em Dokl. Akad. Nauk SSSR} {\bf 98} 527
\\\\
Krylov G and Robnik M 1998 to be published in {\em J. Phys. A: Math. Gen.}
\\\\
Landau L D and Lifshitz E M 1997 {\em Quantum Mechanics} (Oxford:
Butterworth-Heinemann)
\\\\
Lazutkin V F 1981 {\em The Convex Billiard and the Eigenfunctions of the
Laplace Operator} (Leningrad: University Press) (in Russian)
\\\\
Lazutkin V F 1991 {\em KAM Theory and Semiclassical Approximations to
Eigenfunctions} (Heidelberg: Springer)
\\\\
Li Baowen and Robnik M 1994 {\em J. Phys. A: Math. Gen.} {\bf 27} 5509
\\\\
Li Baowen and Robnik M 1995a {\em J. Phys. A: Math. gen.} {\bf 28} 2799
\\\\
Li Baowen and Robnik M 1995b {\em J. Phys. A: Math. gen.} {\bf 28} 4843
\\\\
Li Baowen and Robnik M 1995c Supplement to the Paper (Li and Robnik 1995b),
unpublished 
\\\\
Li Baowen and Robnik M 1996 to be published
\\\\
Markarian R 1993 {\em Nonlinearity} {\bf 6} 819
\\\\
Mather J N 1982 {\em Ergodic theory and Dynamical systems} {\bf 2} 3
\\\\
Mehta M L 1991 {\em Random Matrices} (San Diego: Academic Press)
\\\\
Meyer H D 1985 {\em J. Chem. Phys.} {\bf 84} 3147
\\\\
Moser J 1962 {\em Nachr. Akad. Wiss. G\"ottingen} 1
\\\\
Percival I C 1973 {\em J. Phys. B: At. Mol. Phys.} {\bf 6} L229
\\\\
Poincar\'e H 1993 {\em New Methods of Celestial Mechanics} (American 
Institute of Physics) (English transalation of "M\'ethodes nouvelles de
la m\'echanique c'eleste, 1892-1999) See also the Introduction by
D L Goroff
\\\\
Porter C E and Thomas R G 1956 Phys. Rev. {\bf 104} 483
\\\\
Press W H, Flannery B P, Teukolsky S A and Vetterling W T 1986
{\em Numerical Recipes} (Cambridge: Cambridge University Press) ch 7
\\\\
Prosen T 1994a {\em J. Phys. A: Math. Gen.} {\bf 27} L569
\\\\
Prosen T 1994b {\em Ann. Phys. New York} {\bf 235} 115
\\\\
Prosen T 1995 {\em J. Phys. A: Math. Gen.} {\bf 28} L349
\\\\
Prosen T 1996 {\em Physica D} {\bf 91} 244
\\\\
Prosen T 1997a {\em Phys.Lett.A} {\bf 233} 323
\\\\
Prosen T 1997b {\em Phys.Lett.A} {\bf 233} 332
\\\\ 
Prosen T and Robnik M 1993a {\em J. Phys. A: Math. Gen.} {\bf 26} L319
\\\\
Prosen T and Robnik M 1993b {\em J. Phys. A: Math. Gen.} {\bf 26} 1105
\\\\
Prosen T and Robnik M 1993c {\em J. Phys. A: Math. Gen.} {\bf 26} 2371
\\\\
Prosen T and Robnik M 1993d {\em J. Phys. A: Math. Gen.} {\bf 26} L37
\\\\
Prosen T and Robnik M 1994a {\em J. Phys. A: Math. Gen.} {\bf 27} L459
\\\\
Prosen T and Robnik M 1994b {\em J. Phys. A: Math. Gen.} {\bf 27} 8059
\\\\
Prosen T and Robnik M 1998 {\em J. Phys. A: Math. Gen.} {\bf 31} L345
\\\\
Robnik M 1981 {\em J.Phys.A:Math.Gen.} {\bf 14} 3195
\\\\
Robnik M 1982 {\em J.Physique Colloque C2} {\bf 43} 45
\\\\
Robnik M 1983 {\em J. Phys. A: Math. Gen.} {\bf 16} 3971
\\\\
Robnik M 1984 {\em J. Phys. A: Math. Gen.} {\bf 17} 1049
\\\\
Robnik M 1986 {\em Lecture Notes in Physics} {\bf 263} 120
\\\\
Robnik M 1986 in {\em Nonlinear Phenomena and Chaos} ed S Sarkar
(Bristol: Adam Hilger) 303-330
\\\\
Robnik M 1988 in {\em "Atomic Spectra and Collisions in External Fields"},
eds. K T Taylor, M H Nayfeh and C W Clark, (New York: Plenum) pp265-274
\\\\
Robnik M 1989 Preprint ITP Santa Barbara, unpublished
\\\\
Robnik M 1994 {\em J. Phys. Soc. Japan Suppl. A} {\bf 63} 131
\\\\
Robnik M 1995 {\em Introduction to quantum chaos I} to be published
in the Proc. of the Summer School/Conference in Xanthi, Greece, 1995
\\\\
Robnik M 1997 {\em Open Sys. \& Information Dyn.} {\bf 4} 211
\\\\
Robnik M and Berry M V 1985 {\em J. Phys. A: Math. Gen.} {\bf 18} 1361
\\\\
Robnik M and Berry M V 1986 {\em J. Phys. A: Math. Gen.} {\bf 19} 669
\\\\
Robnik M, Dobnikar J, Rapisarda A, Prosen T and Petkov\v sek M 1997
{\em J.Phys.A:Math.Gen.} {\bf 30} L803
\\\\
Robnik M and Veble G 1998 {\em J. Phys. A: Math. Gen.} {\bf 31} 4669
\\\\
Robnik M, Liu Junxian and Veble G 1998, to be published
\\\\
Seligman T H and Verbaarschot J J M 1985 {\em J. Phys. A: Math. Gen.} {\bf 18} 
2227
\\\\
Shnirelman A L 1979 Uspekhi Matem. Nauk {\bf 29} 181
\\\\
Sinai Ya G 1976 {\em Introduction to Ergodic Theory} (Princeton: Princeton
University Press) p.140
\\\\
Stone A D and Bruus H 1993 {\em Physica B} {\bf 189} 43
\\\\
Stone A D and Bruus H 1994 {\em Surface Science} {\bf 305} 490
\\\\
Strelcyn J.-M. 1991 {\em Colloquium Mathematicum} {\bf LXII} Fasc.2 
331-345
\\\\
Van Kampen N G 1985 in {\em Chaotic Behaviour in Quantum Systems} ed 
Giulio Casati (New York: Plenum) 309
\\\\
Voros A 1979 {\em Lecture Notes in Physics} {\bf 93} 326
\\\\
Weidenm\"uller H A, Verbaarschot J J M and Zirnbauer M 1985 {\em Phys. 
Rep.} {\bf 129} 367-438
\\\\
Wilkinson M 1987 {\em J. Phys. A: Math. Gen.} {\bf 20} 635
\\\\
Wilkinson M 1988 {\em J. Phys. A: Math. Gen.} {\bf 21} 1173

\end{document}